\documentclass[a4paper,12pt,twoside]{article}
\usepackage{epsfig}
\usepackage{amsfonts}
\usepackage{amsmath}
\usepackage{amssymb}
\usepackage{fancyheadings}
\usepackage{theorem}
\usepackage{index}
\usepackage{subfigure}

\usepackage{mathrsfs}

\newindex{default}{idx}{ind}{Author Index}
\def\myhead{
 \newpage
 \pagestyle{fancy}
 \lhead[\thepage]{\it \titles}
 \chead{}
 \rhead[{\it \authors}]{\thepage}
 \lfoot{}
 \cfoot{}
 \rfoot{}
}

\def\titlehead{
	
\vspace*{-2cm} \hspace*{-7mm}

\begin{minipage}[t]{10cm}
Stochastic and Physical Monitoring Systems\\
{\bf  SPMS 2018}\\
June 18 - June 22, 2018\\
Dob\v{r}ichovice, Czech Republic
\end{minipage}\hfill
\begin{minipage}[t]{4cm}
\vspace*{-5pt} \hspace*{5mm}
\mbox{\epsfxsize=33mm\epsffile{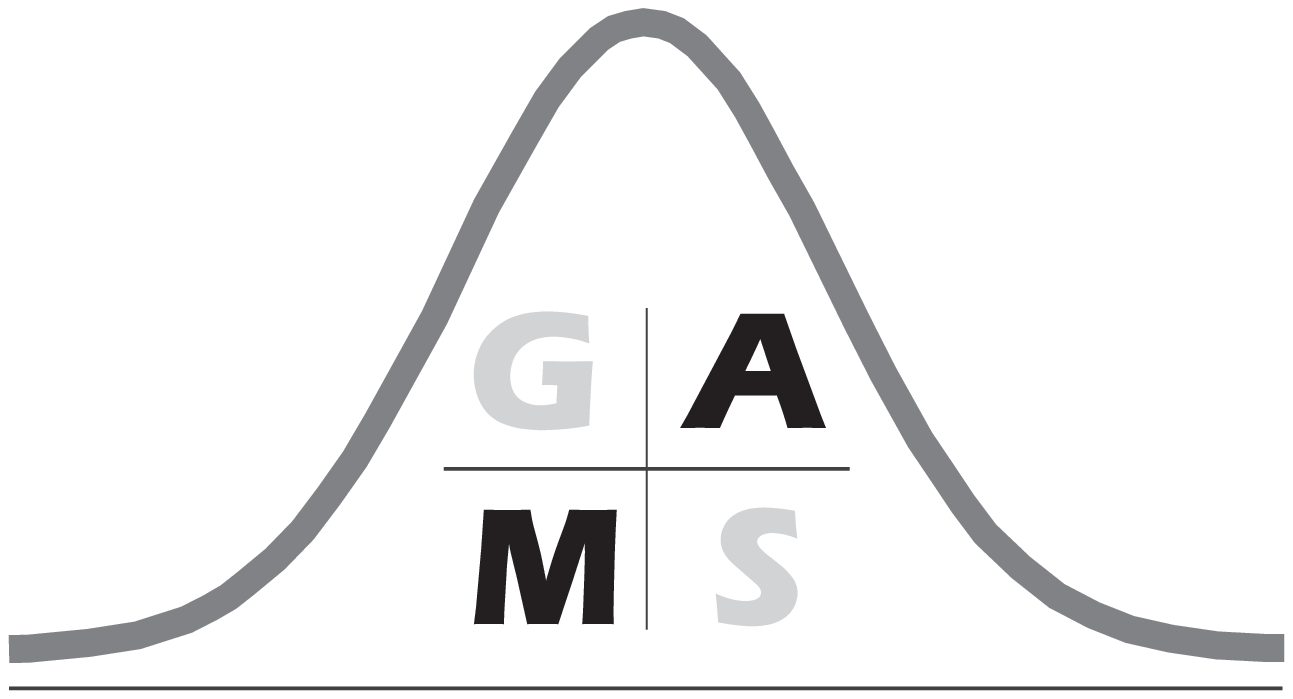}}
\end{minipage}
\rule[-5mm]{\textwidth}{0.7mm}
\bigskip
}

\newcommand{\authorshort}[1]{\def\authors{#1}}
\newcommand{\titleshort}[1]{\def\titles{#1}}

\def\Abstract#1{\small \noindent\textbf{Abstract.} #1 \vskip 1.5em}{\normalsize}

\def\Keywords#1{\small \noindent\textbf{Key words: } #1. \vskip 0.5em}{\normalsize}

\def\Title#1{\vspace*{0.6cm}\noindent {\Large {\bf #1} }
\def\titleall{#1}%
\setcounter{footnote}{0}%
\setcounter{section}{0}
\setcounter{equation}{0}%
\setcounter{figure}{0}%
\setcounter{table}{0}%
\setcounter{theorem}{0}%
\setcounter{remark}{0}%
\setcounter{lemma}{0}%
\setcounter{definition}{0}%
\setcounter{example}{0}%
}
\def\Author#1{\vspace*{1.0cm}\noindent {\large #1 }}
\def\Institution#1{\vspace*{0.4cm}\noindent {#1}}
\def\Email#1{\vspace*{0.3cm}\noindent Email: \texttt{#1}\vspace*{1cm}}

{\theorembodyfont{\upshape}  }
{\theorembodyfont{\upshape}  }

\begin{document}

\newcommand{\IN}{\mathrm{in}}
\newcommand{\OUT}{\mathrm{out}}
\newcommand{\F}{\mathrm{front}}
\newcommand{\B}{\mathrm{rear}}
\newcommand{\BE}{\begin{equation}}
\newcommand{\EE}{\end{equation}}
\renewcommand{\H}{{\mathcal H}}
\newcommand{\e}{\mathtt{e}}
\newcommand{\K}{{\mathcal K}}
\newcommand{\argmin}{\mathrm{argmin}\,}
\newcommand{\Dom}{\mathrm{Dom}}
\newcommand{\Ran}{\mathrm{Ran}}
\newcommand{\argmax}{\mathrm{argmax}\,}
\renewcommand{\d}{\mathtt{d}}
\newcommand{\p}{\partial}
\newcommand{\dx}{\mathtt{d}x}
\newcommand{\dy}{\mathtt{d}y}
\newcommand{\X}{\mathcal{X}}
\newcommand{\Y}{\mathcal{Y}}
\newcommand{\N}{\mathtt{N}}
\newcommand{\M}{\mathtt{M}}
\newcommand{\R}{\mathbb{R}}
\renewcommand{\rho}{\varrho}

\newcommand{\EV}{\mathtt{E}}
\newcommand{\VAR}{\mathtt{VAR}}
\newcommand{\SVAR}{\mathtt{SVAR}}
\newcommand{\SD}{\mathtt{SD}}

\renewcommand{\geq}{\geqslant}
\renewcommand{\leq}{\leqslant}

\myhead

\thispagestyle{plain}
\titlehead


\Title{3s-Unification for Vehicular Headway Modeling}

\titleshort{3s-Unification for Vehicular Headway Modeling}

\Author{Milan Krb\'alek$^{a}$ and Michaela Krb\'alkov\'a$^{b,c}$}
\authorshort{M. Krb\'alek and M. Krb\'alkov\'a}

\Institution{$^{a}$ Department of Mathematics,
Faculty of Nuclear Science and Physical Engineering, Czech Technical University in Prague,
12000 Prague 2, Czech Republic\\\noindent $^{b}$ Department of Physics, Faculty of Science, University of Hradec Kr\'alov\'e, Hradec Kr\'alov\'e, Czech Republic \\ \noindent $^{c}$ Department of Transport Technology and Control, Jan Perner Transport Faculty, University of Pardubice, Pardubice, Czech Republic}

\Email{milan.krbalek@fjfi.cvut.cz}

\Abstract{We explain why \emph{a sampling} (division of data into homogenous sub-samples), \emph{segmentation} (selection of sub-samples belonging to a small sub-area in ID plane -- a segmentation zone), and \emph{scaling} (a linear transformation of random variables representing a standard sub-routine in a general scheme of an unfolding procedure) are necessary parts of any vehicular data investigations. We demonstrate how representative traffic micro-quantities (in an unified representation) are changing with a location of a segmentation zone. It is shown that these changes are non-trivial and correspond fully to some previously-published results. Furthermore, we present a simple mathematical technique for the unification of GIG-distributed random variables. }


\normalsize

\section{Introduction and motivation}

Vehicular dynamics \cite{Traffic Flow Dynamics} is a complex scientific discipline having many interesting components. One of them is \emph{Vehicular Headway Modeling (VHM)} \cite{Li_Headway_Survey} analyzing and predicting changes in a vehicular microstructure forced by external conditions (traffic density, intensity, or global velocity). Typical representatives of vehicular micro-quantities are individual velocities, time/space headways, or time/space clearances. Mathematically, all these quantities represent, in fact, random variables and associated probability densities (for headways and clearances) belonging to a specific family of distributions (see for example \cite{Chodci-2018}). As is well known (see \cite{Krbalek_gas,Traffic_NV,KRB-Kybernetika,My_Multiheadways,Red-intervals}), parameters of these distributions evolve rapidly over time and are therefore markedly dependent on actual values of density $\rho,$ intensity $I,$ and global velocity $V$. It excludes the possibility of applying standard statistical approaches to entire data structures and, on contrary, it enforces the use of more complex procedures applied to partial data samples. The description of these methods is the main goal set for this article.

\section{Typical sets of empirical traffic data} \label{sec:Typical-sets}

There exist many technologies suitable for vehicle-by-vehicle measurements. They are typically divided into two categories: \emph{intrusive} (induction loops, piezo-electric cables, active infrared sensors, etc) and \emph{non-intrusive} (passive infrared sensors, detection drones, ultrasonic sensors, video image processing, etc). However, if one aims to deal with Big Data sets the most of these methods are inappropriate since they provide a limited amount of data only. In contrast, commonly available magnetic induction double-loop detectors (as an example of intrusive measurement techniques) do not suffer from this inefficiency. Therefore, in the rest of the text we consider data gauged by magnetic loops or by similar technology.

\begin{figure}[!h]
  \vspace{0.2cm}
  \centering
   {\epsfig{file = 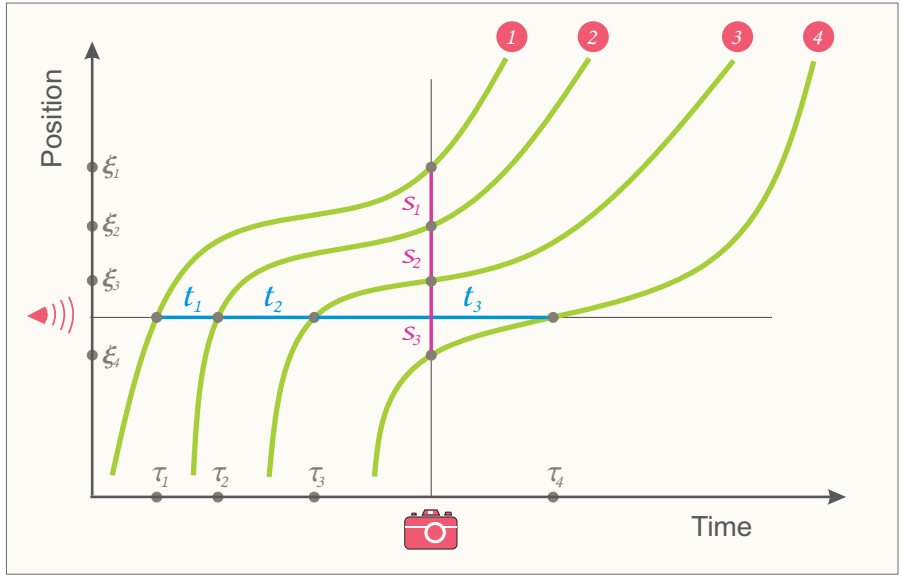, height=3in}}
  \parbox{12cm}{\caption{\footnotesize{An illustration of time and space headways on the space-time plot. Each curve symbolizes a trajectory of a front car-bumper.\label{fig:trajektorie}}}}
  \vspace{-0.1cm}
\end{figure}

Typical attribute of those data is that they have been measured at the same time-fixed location (a so-called \emph{detector line}). Considering now one-lane traffic flow (without loss of generality) it is easy to comprehend that characteristic outputs  of a traffic measurement look like
\BE T^{(\IN)}=\bigl\{\tau^{(\IN)}_k \in\R ~|~  k\in\hat{\N} ~\wedge~ \tau^{(\IN)}_{k-1} < \tau^{(\IN)}_{k}  ~\wedge~ \tau^{(\IN)}_0:=0\bigr\}, \label{casIN}\EE
\BE T^{(\OUT)}=\bigl\{\tau^{(\OUT)}_k \in\R ~|~  k\in\hat{\N} ~\wedge~ \tau^{(\IN)}_{k-1} <  \tau^{(\OUT)}_{k-1} \leq \tau^{(\IN)}_{k}< \tau^{(\OUT)}_{k}  ~\wedge~ \tau^{(\OUT)}_0:=0\bigr\}, \label{casOUT}\EE
\BE \Upsilon=\bigl\{v_k \in\R^+_0 ~|~  k\in\hat{\N}\bigr\}, \label{velo}\EE
\BE \Lambda=\bigl\{\ell_k \in\R^+ ~|~  k\in\hat{\N}\bigr\}, \label{delky-aut}\EE
where $T^{(\IN)}$ and $T^{(\OUT)}$ collect instants $\tau_k$ when a front/rear bumper of a $k$th car has intersected a detector line and $\Upsilon$ and $\Lambda$ collects individual velocities $v_k$ and lengths $\ell_k$ of cars, respectively.

If loop measurements are accompanied by image processing technology then additional data sets
\BE \Xi^{(\F)}=\bigl\{\xi^{(\F)}_k \in\R ~|~  k\in\hat{\N}~\wedge~   \xi^{(\F)}_{k} < \xi^{(\F)}_{k-1} ~\wedge~ \xi^{(\F)}_0:=+\infty \bigr\}, \label{xi-F}\EE
\BE \Xi^{(\B)}=\bigl\{\xi^{(\B)}_k \in\R ~|~  k\in\hat{\N}~\wedge~ \xi^{(\B)}_{k}       < \xi^{(\F)}_{k} \leq \xi^{(\B)}_{k-1} < \xi^{(\F)}_{k-1} ~\wedge~ \xi^{(\B)}_0:=+\infty  \bigr\}, \label{xi-B}\EE
(collecting positions of front/rear bumpers at the fixed time) are to disposal. However, such a doubled measurements are very rare, which results in the fact that locations $\xi_k$ (in contrast to instants of time $\tau_k$) belong to  indirectly determined traffic quantities. Therefore, we refer instants of time $\tau_k,$ velocities $v_k,$ and lengths $\ell_k$ as \emph{primary quantities,} whereas locations $\xi_k$ are referred to as \emph{secondary quantities.}

\begin{figure}[!h]
  \vspace{0.2cm}
  \centering
   {\epsfig{file = 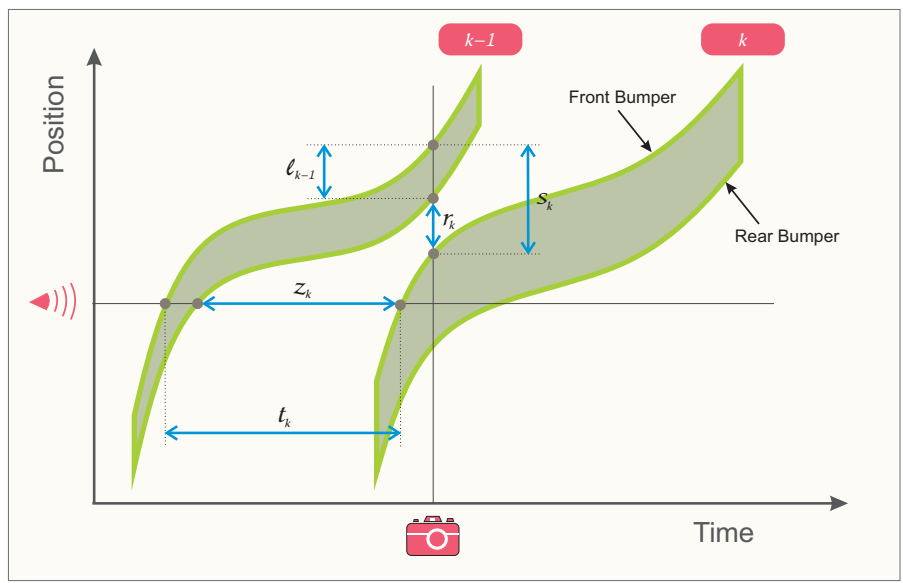, height=3in}}
  \parbox{12cm}{\caption{\footnotesize{Individual vehicular characteristics obtained by cross-section detectors, e.g. induction double-loop detectors, $(z_k,t_k)$ or by image processing technology  $(r_k,s_k).$\label{fig:trajektorie-2}}}}
  \vspace{-0.1cm}
\end{figure}

\section{Random variables in VHM}

Less formally (see for example \cite{Li_Headway_Survey}), in Vehicular Headway Modeling (VHM) a time headway is usually defined as the time between two successive vehicles as they pass a fixed point on the same lane. Analogously, a space headway is understood as distance between the same common feature (e.g., front/rear bumper) of two subsequent vehicles measured at a fixed time. Except headways, VHM often works with vehicular clearances. Time clearance represents the time between two following events: 1) rear bumper of a previous car intersects a detector line; 2) front bumper of a reference car intersects a detector line. Distance clearance is, in analogy, defined as a space gap between vehicles measured at a fixed time. Illustratively, all these quantities are visualized in figures \ref{fig:trajektorie} and \ref{fig:trajektorie-2}. More formally, they will be defined in the following subsection.

\subsection{Standard typology of empirical headways} \label{sec:headway-typology}

Now, knowing typical outputs of traffic measurements one can define empirical values of \emph{time headways/clearances} using respective formulas
$$t_k:= \tau^{(\IN)}_{k} - \tau^{(\IN)}_{k-1}, \quad z_k:= \tau^{(\IN)}_{k} - \tau^{(\OUT)}_{k-1}.$$
Distance headways are then defined by
$$s_k:= \xi^{(\F)}_{k-1} - \xi^{(\F)}_{k}$$
and distance clearances read
$$r_k:= \xi^{(\B)}_{k-1} - \xi^{(\F)}_{k}.$$
Note that equalities $s_k=v_kt_k,$ $r_k=v_kz_k$ are valid only provided that an individual velocity $v_k$ is constant during time interval $[\tau^{(\IN)}_{k-1},\tau^{(\IN)}_{k}],$ which is less probable assumption, especially when $\tau^{(\IN)}_{k} \gg \tau^{(\IN)}_{k-1}.$ It means that loop measurements (without additional processing of video/photography) generate time clearances and time headways as primary variables, whereas distance headways/clearances (taking into account the method of data collection) represent secondary variables and their approximative values $s_k \approx v_kt_k,$ $r_k \approx v_kz_k$ can be therefore burdened by systematic errors. Moreover, one finds
$$r_k=\xi^{(\B)}_{k-1} - \xi^{(\F)}_{k-1}+ \xi^{(\F)}_{k-1}- \xi^{(\F)}_{k} = s_k-\ell_{k-1}.$$

\subsection{Stochastic representation of vehicular headways}

Mathematically, all the quantities from the subsection \ref{sec:headway-typology}  represent non-negative continuous random variables and are therefore characterized by a standard statistical description using associated probability densities and distribution functions (cumulated probability densities). Consider now a sequence $(\X_k)_{k=1}^\N$ of random variables of the same type (clearance, headway). From a statistical viewpoint the empirical headways $x_k$ represent individual realizations of random variables $\X_k$ and one can model respective distributions by standard statistical routines. Generally accepted premise in VHM says that $\X_1,\X_2,\ldots ,\X_\N$ are identically distributed provided that one analyzes homogeneous flows, where macroscopic quantities \emph{(state variables)} are steady in time and a rate of long vehicles is low. To a certain degree of simplification it used to be sometimes speculated that $\X_1,\X_2,\ldots ,\X_\N$ are i.i.d. (e.g. \cite{Cowan}), which is definitely not a reasonable hypothesis. However, in many analytical studies this represent a useful (and simplifying) assumption. In fact, correlations among headways are significant \cite{Szabova} and reveal a more complex interaction rules among drivers. This is, unfortunately, beyond the scope of this article.

\section{Macroscopic characteristics of traffic}

Usually, three basic phase variables (density, intensity and mean speed), describing a macroscopic constellation of each vehicular ensemble, are defined somewhat vaguely (see pages 15-17 in \cite{Traffic Flow Dynamics}). Intensity $I(x,\tau) = \Delta N/\Delta \tau$ is understood as the number of vehicles $\Delta N$ passing a given cross-section at locations $x$ within a time interval $\Delta \tau,$  whereas density $\varrho(x,\tau) = \Delta N/\Delta x $ represents the number of vehicles $\Delta N$ lying in an interval $[x,x+\Delta x]$ during a fixed time $\tau.$ Especially these two definitions are difficult to interpret since number of vehicles is a discrete quantity not allowing standard differentiations. Therefore, it is necessary to go to a mathematically correct formulation.

\subsection{Theoretical definitions}

Consider a population of $\M$ succeeding (dimensionless) vehicles located in a fixed time $\tau$ at locations $\alpha_1(\tau) > \alpha_2(\tau) > \ldots > \alpha_\M(\tau).$ Let $\wp(x|\sigma)\in\mathscr{C}^2(\R)$ be an arbitrary probability density having $0$ as an expected value and $\sigma$ as a shape-parameter. Typical representatives of generating densities are summarized in Appendix \ref{subsec:APP-01}. Then, after a selection of a suitable generating density, one can define \emph{smoothed number of particles} as
\BE N(\xi,\tau):=\int_{-\infty}^\xi \sum_{k=1}^\M \wp(y-\alpha_k(\tau)|\sigma)\,\dy. \label{SNP}\EE
Using such a function the \emph{density} and \emph{intensity} can be defined as
\BE \varrho(\xi,\tau):= \frac{\p N(\xi,\tau)}{\p \xi}, \quad I(\xi,\tau):= -\frac{\p N(\xi,\tau)}{\p \tau},\EE
respectively. Moreover, both can be simplified into following forms:
\BE \varrho(\xi,\tau)= \sum_{k=1}^\M \wp(\xi-\alpha_k(\tau)|\sigma), \EE
\BE I(\xi,\tau)= \sum_{k=1}^\M \frac{\d \alpha_k(\tau)}{\d \tau} \wp(\xi-\alpha_k(\tau)|\sigma) = \sum_{k=1}^\M v_k(\tau) \wp(\xi-\alpha_k(\tau)|\sigma). \EE
Owing to the prerequisite $\wp(\xi|\sigma)\in\mathscr{C}^2(\R)$ we have
\BE \frac{\p \varrho}{\p \tau}=\frac{\p^2 N(\xi,\tau)}{\p \xi \p \tau}=\frac{\p^2 N(\xi,\tau)}{\p \tau \p \xi} = - \frac{\p I}{\p \xi}, \label{continuity-eq} \EE
which corresponds to the well-known \emph{equation of continuity,} here interpreted as the law of conservation for the number of vehicles. If we restrict our considerations to homogeneous flows only, where $v_k(\tau)=V(\tau),$ then we obtain a hydrodynamic equality
\BE I(\xi,\tau)=V(\tau)\varrho(\xi,\tau), \label{hydro-eq}\EE
that usually serves itself as a generally accepted approximation for a fundamental relationship among three basic traffic state variables. However, always it is necessary to keep in mind that equation $I(\xi,\tau)=V(\xi,\tau)\varrho(\xi,\tau),$ contrary to equation (\ref{continuity-eq}), does not represent a universally valid formula.

\subsection{Fundamental phase relations in vehicular traffic}

\begin{figure}[!h]
  \vspace{0.2cm}
  \centering
   {\epsfig{file = 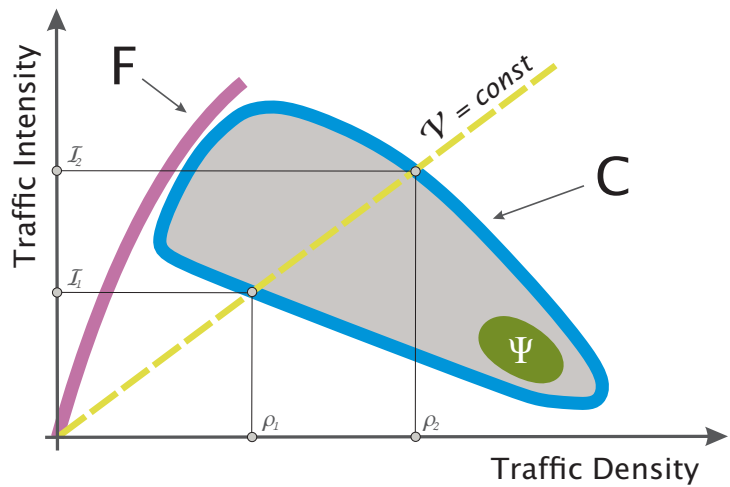, height=2.5in}}
  \parbox{12cm}{\caption{\footnotesize{Schematic visualization of fundamental relations. Curve $F$ represents \emph{free-flow traffic,} whereas blue area $C$ corresponds to \emph{congested traffic,} where one can reveal (using an analysis of real-time trajectories) two sub-regimes: \emph{synchronized flow} or \emph{wide moving jam} (see \cite{Review-Kerner} for details). Note that, even a velocity of cars is almost constant (i.e. near a yellow line), set of admissible densities covers the interval $[\varrho_1,\varrho_2]$ and set of admissible intensities covers the interval $[I_1,I_2].$ Such behavior is in a direct contradiction with theories involving the fundamental diagram of traffic flow, which suppose a one-dimensional relationship between density and intensity. Green surface represents a selected segment (subset) of ID plane (a segmentation zone). \label{fig:id-plane} }}}
  \vspace{-0.1cm}
\end{figure}

\begin{figure}[!h]
  \vspace{0.2cm}
  \centering
   {\epsfig{file = 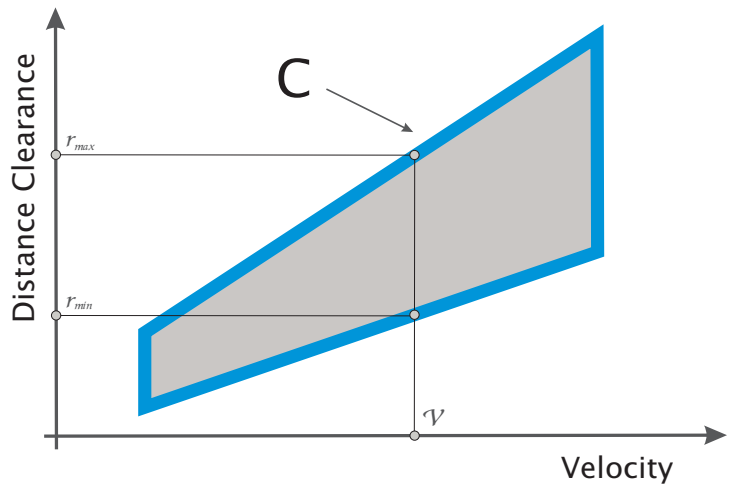, height=2.5in}}
  \parbox{12cm}{\caption{\footnotesize{Schematic visualization of Kerner's hypothesis. Steady states of a homogeneous synchronized flow are spread on two-dimensional region. \label{fig:kernerovo} }}}
  \vspace{-0.1cm}
\end{figure}

In physics of traffic, \emph{fundamental diagrams} are understood as graphical visualizations of relationship among the three basic traffic macro-quantities (phase variables). In earlier works (see review \cite{Review-Helbing}) authors started from the proposition that traffic intensity $I$ and density $\varrho$ are interconnected by a certain functional dependence $I=I(\varrho),$ whose graphical interpretation has a shape of a curve $H(I,\varrho)=0$ lying within ID plane. Analogously, it was speculated that $V=V(\varrho)$ is a function and its graph is a curve. Kerner's three-phase theory \cite{Review-Kerner} and his hypothesis about two-dimensional states of traffic flow, however, disproved this fact by the following claim:

\emph{Consider a homogeneous synchronized flow, that is here understood as a hypothetical state of synchronized flow of identical vehicles and identical drivers in which all vehicles move with the same time-independent speed (see the yellow curve in figure \ref{fig:id-plane}) and have the same space gaps. Surprisingly, these homogeneous synchronized flows can  occur anywhere in some two-dimensional sub-region of the ID plane. It means that at a given speed in congested flow, a driver does not use some unique gap. Quite the contrary, a driver can make an arbitrary choice for the space clearance to a preceding vehicle (see figure \ref{fig:kernerovo}), which leads to the fact that (although a speed of the vehicles is constant) the set of admissible states (described by intensity and density) is infinite. Real-road traffic states with the same average speed therefore cover, in fact, a certain two-dimensional area lying inside the ID plane.}

Mathematically, it means that, instead of a functional relationship $I=I(\varrho),$ we need to switch to a mathematical description using a term of \emph{a binary relation,} which, generally speaking, represents an arbitrary set of ordered pairs. Thus, for vehicular applications \emph{a binary ID relation} $\Omega_{\mathtt{ID}}$ is defined a set of all pairs $(\varrho,I)\in (0,+\infty)^2,$ which can be gauged for a given vehicular stream. Then \emph{the domain} of $\Omega_{\mathtt{ID}}$ is the set of all $\varrho$ such that $(\varrho,I)\in\Omega_{\mathtt{ID}}$ for at least one $I.$ \emph{The range} is the set of all $I,$ for which exists at least one $\varrho$ such that $(\varrho,I)\in\Omega_{\mathtt{ID}}.$ Thus, $\Omega_{\mathtt{ID}} \subset \Dom(\Omega_{\mathtt{ID}}) \times \Ran(\Omega_{\mathtt{ID}}).$ In a similar way, the \emph{a binary VD relation} $\Omega_{\mathtt{VD}}$ between mean speed and density is introduced.

\subsection{Empirical extractions of fundamental binary relations}\label{subsec:EE}

The way to get a graphical shape of the fundamental relations \emph{(a phase diagram)} from empirical data is not difficult. For a given sample
\BE \Xi=\left\{(\tau^{(\IN)}_k,\tau^{(\OUT)}_k,v_k,\ell_k)\in T^{(\IN)} \times T^{(\OUT)} \times \Upsilon \times \Lambda:~k=1,2,\ldots,\M\right\}. \label{eq:data-samples} \EE
of $\M$ succeeding cars one can calculate the local intensity $I=\M/(\tau_{\M}^{(\OUT)}-\tau_1^{(\IN)})$ and the local mean speed $V=\M^{-1}\sum_{k=1}^\M v_{k}.$ Besides, the expression $\varrho=I/V$ used to be (in accordance with the relationship (\ref{hydro-eq})) usually accepted as a plausible approximation for the local density (according to \cite{My_Multiheadways,Review-Helbing}). Then, applying the same computational routine for all available data samples $i=1,2,\ldots,m$ we acquire empirical binary relations $\Omega_{\mathtt{ID}}=\{(\varrho_i,I_i):~i=1,2,\ldots,m\}$ and $\Omega_{\mathtt{VD}}=\{(\varrho_i,V_i):~i=1,2,\ldots,m\},$ whose graphs can be used for synoptic visualizations of macroscopic properties of traffic flow.

\section{3s-Unification}

Specific and well-known signs of traffic are (see \cite{Review-Helbing}): a strong non-linearity of congested states, chaotic evolution of state variables, repetitive sharp increases of density, and a propagation of kinematic
waves in a direction opposite to vectors of vehicular velocities. All these effects cause that larger samples of succeeding vehicles show a significant non-homogeneities. Their microstructure is therefore also non-homogeneous which results in the fact that associated probability distributions are not one-component ones and produce, in contrast, courses typical for mixed systems composed from several different distributions.

To obtain homogeneous characteristics one has to apply the following 3s-unification procedure that prevents an undesirable mixing of states with different statistical properties (like resistivity, stochastic rigidity, or compressibility), different vehicular properties (headways, clearances, velocities) and psychological properties (vigilance of drivers, reaction times, decision-making strain).

\subsection{Sampling}

The first sub-routine in a three-phase unification procedure is \emph{the sampling,} i.e. division of data into homogeneous samples of several neighboring cars.

Consider the data sets (\ref{casIN})--(\ref{delky-aut}), a sampling size $\M,$ and number of samples $m.$ Without loss of generality, we assume that $m\M=\N.$ For each sample $i\in\{1,2,\ldots,m\}$ we denote $G_i=\{(i-1)\M+1,(i-1)\M+2,\ldots,i\M\}$ a respective index set and extract relations $\Omega_{\mathtt{ID}}^{(i)}$ and $\Omega_{\mathtt{VD}}^{(i)}$ as introduced in subsection \ref{subsec:EE}. Moreover, we define sample-adjoint sets of individual headways (time and spatial) $T_i:=\{t_k:k\in G_i\}$ and $S_i:=\{s_k:k\in G_i\},$ sets of individual clearances (time and spatial) $Z_i:=\{z_k:k\in G_i\}$ and $R_i:=\{r_k:k\in G_i\},$ and sets of velocities $\Upsilon_i:=\{v_k:k\in G_i\}$ and lengths $\Lambda_i:=\{\ell_k:k\in G_i\}$. From this approach it follows that the $i$th sample is described by the sample-adjoint values $\varrho_i,I_i,V_i$ and by the random sets
\BE T_i,S_i,Z_i,R_i,\Upsilon_i,\Lambda_i. \label{random-sets} \EE

\subsection{Scaling}

As understandable, the mean values $\langle T_i \rangle,\langle S_i \rangle,\ldots,\langle \Lambda_i \rangle$ of the random data sets (\ref{random-sets}) can be easily enumerated by means of the values  $\varrho_i,I_i,V_i.$ Indeed, it holds
\BE \langle T_i \rangle = 1/I_i; \quad \langle S_i \rangle = 1/\varrho_i; \quad \langle \Upsilon_i \rangle = V_i \EE
and $\langle R_i \rangle = \langle S_i \rangle - \langle \Lambda_i \rangle.$ It means that for traffic micro-quantities the statistical characteristics of  the first order are hidden in relations $\Omega_{\mathtt{ID}}$ and $\Omega_{\mathtt{VD}}.$ Therefore, a scaling of individual micro-quantities  represents no loss of information. Above that, the standard unfolding procedure (usually applied in many statistical studies aiming to reveal a non-trivial stochastic universality like in Random Matrix Theory \cite{Mehta,Hobza-RMT}) includes a scaling procedure as its integrated part. For example, by transition to the same expected value, two different random variables can be identified as identically distributed or as variables belonging to the same one-parametric distribution family where the one and only parameter rules a respective variance.

Mathematically, a scaling is understood as a special variant of a general \emph{affine transformation} mapping a random variable $\X$ (here understood as a non-negative and continuous random variable)  into new random variable  $\Y=a\X+\mu,$ where $a> 0$ and $\mu=0.$ As simply follows from elementary chapters of theory of probability, if $g(x)$ is a probability density associated with $\X$ then
\BE h(y)=\frac{1}{a}g\left(\frac{y-\mu}{a}\right) \EE
is a density associated with $\Y=a\X+\mu.$ Thus,
\BE \EV(\Y)=a\EV(\X)+\mu, \quad \VAR(\Y)=a^2\VAR(\X). \label{new-EV-VAR} \EE

Let $\X$ be a non-negative and continuous variable with expected value $\EV(\X)>0.$ Then we define \emph{a scaling (scaling transformation)} as the affine transformation $\Y=a\X,$ for which $a=1/\EV(\X).$ Equalities (\ref{new-EV-VAR}) are resulting in $\EV(\Y)=1$ and $\VAR(\Y)=\VAR(\X)/\EV^2(\X).$ It means that the scaling transformation maps all random variables into variables having the unit expected value. For empirical/experimental data the scaling should be applied as follows. Traffic micro-quantities (clearances or headways) are converted to associate scaled alternatives. Such a conversion is here demonstrated on the example of the sample-adjoint set $Z_i:=\{z_k:k\in G_i\}$ of time clearances. Associate set $Y_i:=\{y_k:k\in G_i\}$ of scaled clearances is calculated using a definition
\BE y_k=\frac{z_k\cdot\M}{\sum_{k \in G_i} z_k} = \frac{z_k}{\langle Z_i \rangle}; \quad (k\in G_i) \EE
which ensures that $\langle Y_i \rangle = 1.$ In analogy, we define scaled spatial clearances by
\BE x_k=\frac{r_k\cdot\M}{\sum_{k \in G_i} r_k}  = \frac{r_k}{\langle R_i \rangle}; \quad (k\in G_i). \EE

\subsection{Segmentation}

The final step of a three-phase unification procedure is forced by the fact that most of traffic variables are significantly changing if the state variables vary. It means that random variable characteristics of the first, second, third,  and fourth order (average, variance, skewness, and kurtosis, respectively) strongly depend on actual values of traffic macro-quantities. Naturally, a mixing of different traffic states (i.e. states with different values of phase variables) is significantly undesirable. For this reason, one has to analyze data from small sub-area of a phase diagram only. To be specific, denoting $\Psi$ an arbitrary subset (a small, typically) of the $\Dom(\Omega_{\mathtt{ID}}) \times \Ran(\Omega_{\mathtt{ID}})$ (see the green surface in figure \ref{fig:id-plane} as an example) the \emph{segmentation procedure} selects those samples having phase relations belonging to $\Psi.$ Therefore, we introduce a $\Psi-$adjoint index set (referred to as \emph{segmented index set})
\BE F_\Psi := \left\{i=1,2,\ldots,m:\, (\varrho_i,I_i)\in \Psi\right\} =  \{i=1,2,\ldots,m:\, \Omega_{\mathtt{ID}}^{(i)}\subset \Psi\}.\EE
Then, statistical analysis intended is performed separately for scaled micro-quantities
\BE Y_\Psi\times X_\Psi \times \Upsilon_\Psi:= \{(y_k,x_k,v_k)\in Y_i \times X_i \times \Upsilon_i:\, i\in F_\Psi \,\wedge\, k\in G_i\} \label{svinisce} \EE
extracted from \emph{a phase segment $\Psi.$}

\subsection{Unification procedure: less formally}

The entire data file is divided into small samples of $\M$ successive vehicles \emph{(a sampling stage).} Then, in all these samples, associate random variables (headways, clearances) are scaled so that in every sample the mean value is equal to one \emph{(a scaling stage).} For each sample the phase variables (density, intensity, and mean speed) are calculated. A small phase segment $\Psi$ in ID plane is chosen. Samples whose phase variables lie outside this segment are eliminated from further data processing. It means that statistical distributions of headways/clearances/individual velocities are analyzed (and estimated) for almost homogeneous data belonging to a small sub-region located within a graph of the fundamental phase relation $\Omega_{\mathtt{ID}},$ or $\Omega_{\mathtt{VD}},$ alternatively \emph{(a segmentation stage).}

\section{Statistical analysis of unified traffic data}

In this section we analyze 3s-unified traffic sequences (\ref{svinisce}) obtained from induction double-loop measurements performed at the Expressway R1 (also called the Prague Ring) in
Prague, the Czech Republic.

\begin{figure}[!h]
  \vspace{0.2cm}
  \centering
   {\epsfig{file = 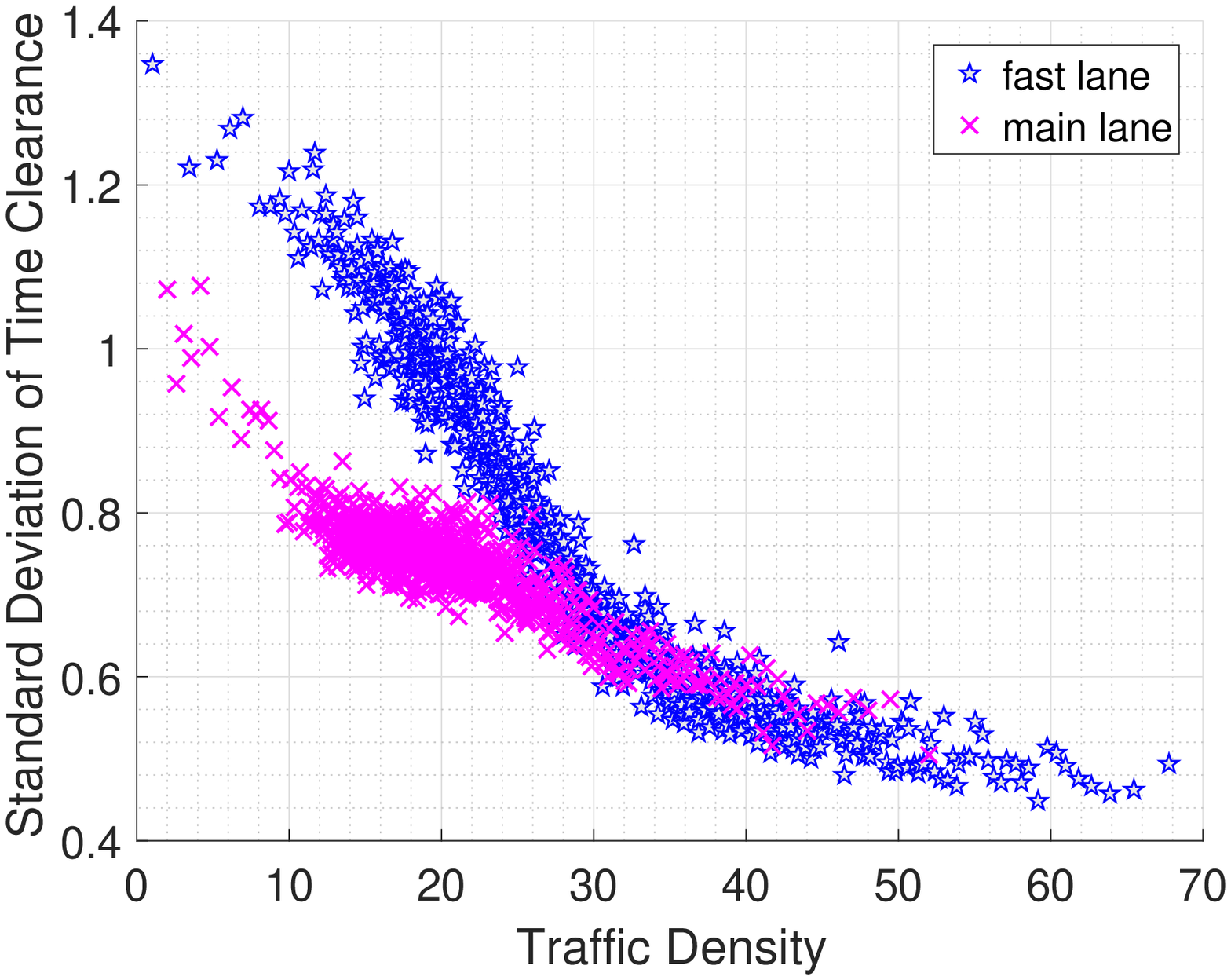, height=2.5in}}
  \parbox{12cm}{\caption{\footnotesize{Standard deviation $\SD(Y_\Psi)$ of time clearance as a function of traffic density. \label{fig:StDev clearance} }}}
  \vspace{-0.1cm}
\end{figure}

\begin{figure}[!h]
  \vspace{0.2cm}
  \centering
   {\epsfig{file = 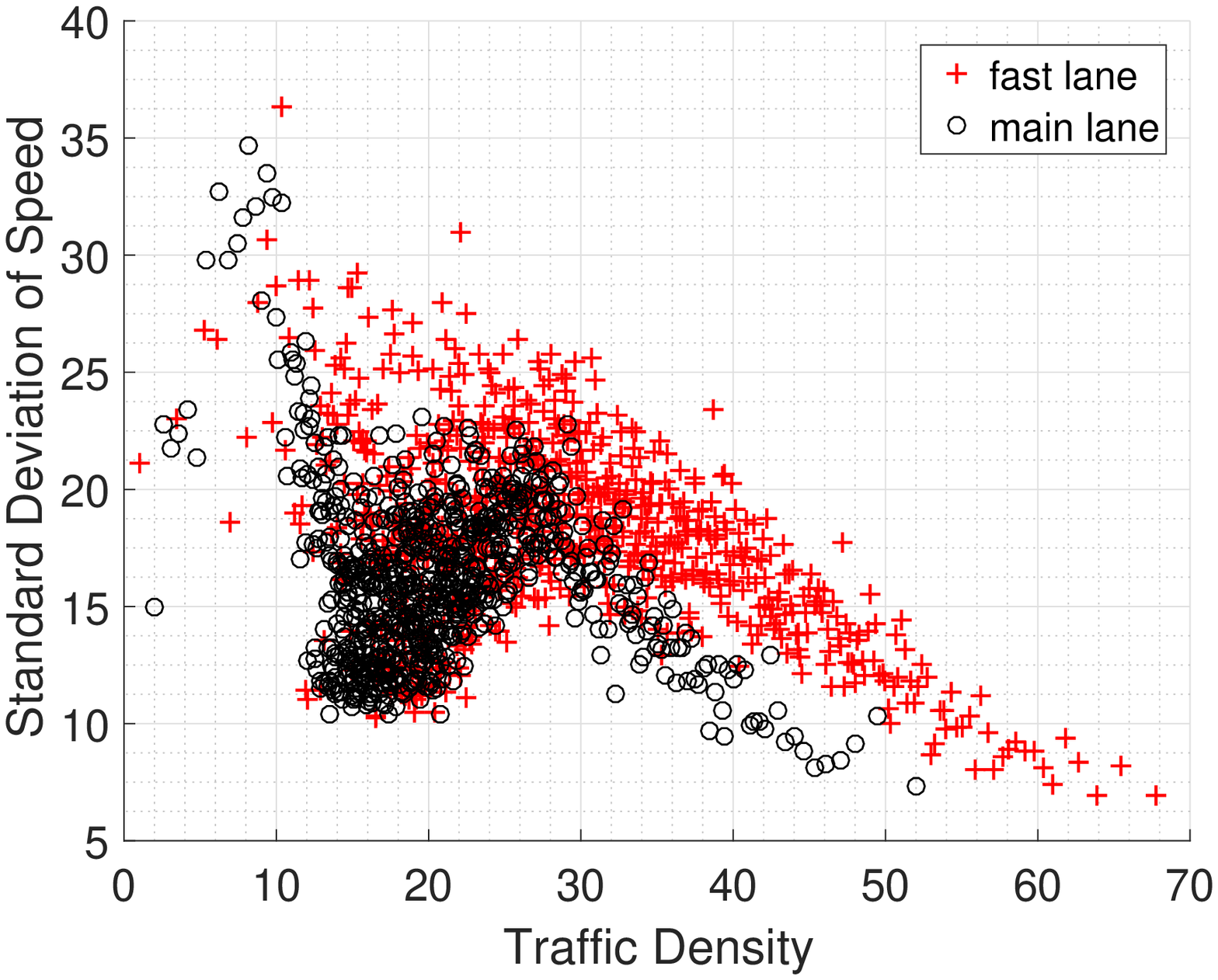, height=2.5in}}
  \parbox{12cm}{\caption{\footnotesize{Standard deviation $\SD(V_\Psi)$ of speed as a function of traffic density. \label{fig:StDev speed} }}}
  \vspace{-0.1cm}
\end{figure}

In order to illustrate typical outputs of statistical analysis applied to 3s-unified data we carry out the following procedure. For each sample $Y_i$ (and $V_i$) of $M=50$ succeeding values of scaled time clearances (and velocities)  the pair $(\rho_i,I_i)$ is calculated.
Consecutively, these samples are arranged by density in ascending order so that $\rho_i < \rho_{i+1}.$ Then we split the ordered samples into $779$ sets (each having $m=30$ samples, i.e. $1500$ individual data).  It means that $\Psi_k=[ \rho_k,\rho_k+\Delta_k) \times (0,+\infty),$ where $\Delta_k$ is determined so that $|Y_{\Psi_k}|=|V_{\Psi_k}|=1500$ and  $\rho_{k+1}=\rho_k+\Delta_k.$ This allows to enumerate (for each set) the standard deviation $\SD(Y_{\Psi_k}),$ $\SD(V_{\Psi_k})$ of unified time clearances and non-scaled velocities. Results of the analysis are depicted in figures \ref{fig:StDev clearance} and \ref{fig:StDev speed}.

In the figure \ref{fig:StDev clearance} we detect an extremely surprising behavior of standard deviation $\SD(Y_\Psi).$ Although the theory of one-dimensional vehicular gases (driven by repulsive potentials and influenced by non-zero level of stochastic resistivity - see \cite{Krbalek_gas}) asserts that scaled inter-vehicular gaps can show variances smaller than one (see \cite{Traffic_NV}), the standard deviation $\SD(Y_\Psi)$ enumerated for vehicles moving in an overtaking lane of a two-lane freeway violates this theoretical restriction. It means that a simulation concept (using thermodynamical approaches applied to particle ensembles with a strict repulsion among particles) is not suitable for simulations of movements in an overtaking lane (for low traffic densities). Furthermore, differences between stars and crosses in figure \ref{fig:StDev clearance} clearly confirms the general opinion that movements of vehicles in fast lane, in contrast to vehicles moving in main lane, are driven by different rules/forces/potentials. This knowledge represents a very interesting open problem in physics of traffic.

\section{{Conclusions}}
\label{sec:conclusion}

We introduce the 3s-unification procedure whose application is eligible if analyzing any vehicle-by-vehicle data by means of statistical methodology. We explain why all the three stages of this unifications (sampling, segmentation, and scaling) are requisite for correct understanding of statistical properties in vehicular microstructure. We mathematically formalize a structure and description of typical traffic data and provide a theoretically correct alternative for a definition of elementary macroscopic quantities in physics of traffic and precise/approximate relations among them.

Finally, we show how sample standard deviations $\SD(Y_\Psi),$ $\SD(V_\Psi)$ of unified time clearances $Y_\Psi$ and non-scaled velocities $V_\Psi,$ respectively, are changing with a location of a phase segment $\Psi.$ Moreover, a thorough statistical analysis of standard deviations reveals a new open problem in the field VHM.

\subsubsection*{Acknowledgements}
Research presented in this work has been supported by the Grant SGS18/\-188/\-OHK4/\-3T/\-14 provided by the Ministry of Education, Youth, and Sports of the Czech Republic (M\v SMT \v CR). The authors would also like to thank The Road and Motorway Directorate of the Czech Republic (\v Reditelstv\' i silnic a d\'alnic \v CR) for providing traffic data analyzed in this paper.

\section*{Appendix}

\subsection{Generating densities} \label{subsec:APP-01}

\begin{figure}[!h]
  \vspace{0.2cm}
  \centering
   {\epsfig{file = 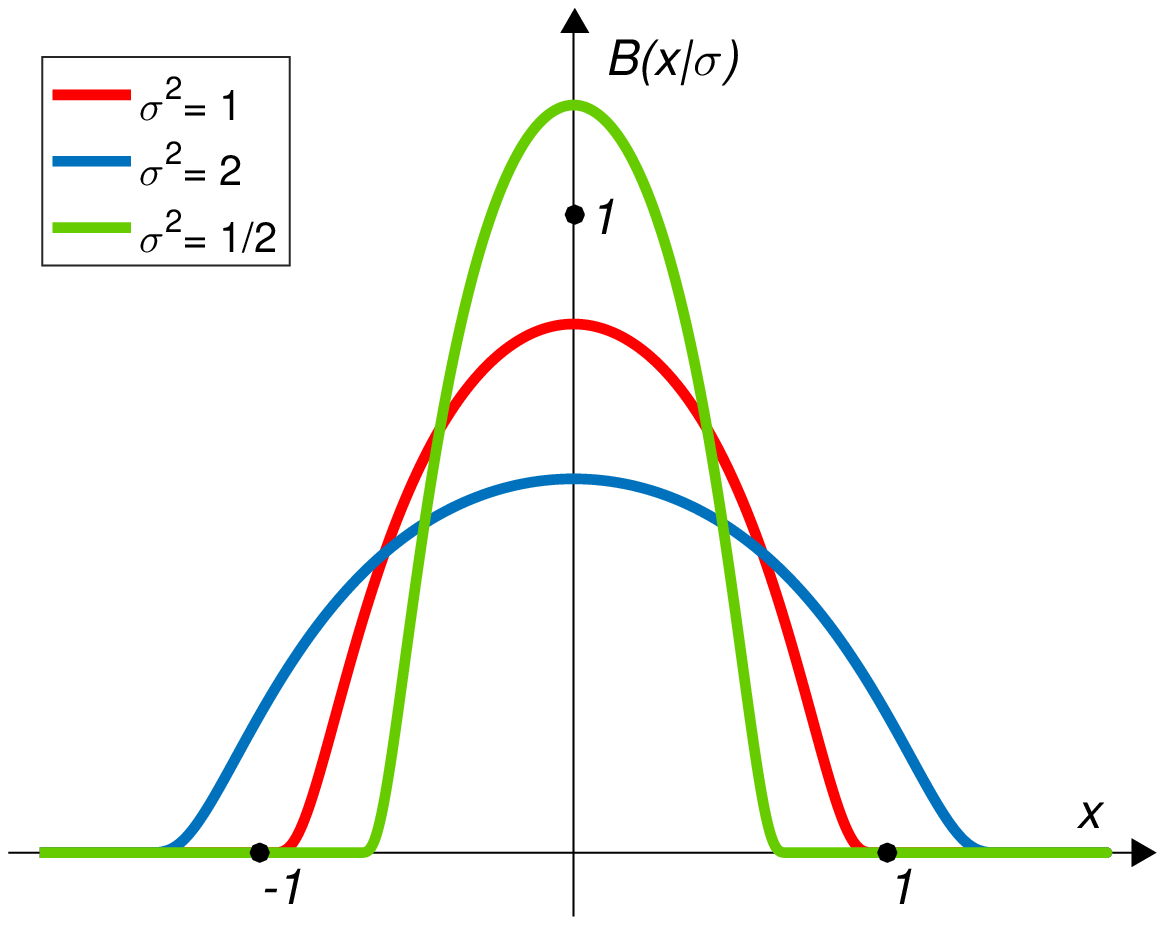, height=2.5in}}
  \parbox{12cm}{\caption{\footnotesize{Borsalino functions. Center is $\alpha=0$ and the shape-parameter is $\sigma\in\{1,\sqrt{2},\sqrt{2}/2\}.$ \label{fig:borsalino} }}}
  \vspace{-0.1cm}
\end{figure}

To build the smoothed number of particles (\ref{SNP}) it is necessary to select a suitable generating density. For these purposes it can be used the most usual generator, namely probability density
$$\wp(x|\sigma)=\frac{1}{\sigma \sqrt{2\pi}} \e^{-\frac{x^2}{2\sigma^2}}$$
of normal-distributed random variable. This Gaussian generator, however, has not a bounded support, which is somewhat unfavorable in vehicular applications. Instead, we can use more appropriate function family derived from the probability density
\BE \omega(x)=
\frac{1}{\mathtt{Z}}\Theta(1-|x|)\e^{\frac{1}{x^2-1}}\label{sapocka}\EE
where $\mathtt{Z} \approx 0.4439940$ and $\Theta(x)$ stands for the Heaviside unit-step function. Then for $\sigma>0$ the function
$$ B(x|\sigma):= \frac{1}{\sigma} \omega\left(\frac{x}{\sigma}\right)$$
is called \emph{Borsalino function} (see figure \ref{fig:borsalino} and a note \cite{Borsalino}) centered to the point $x=0$ and shaped by the parameter $\sigma.$ In general, a generating function $\wp(x|\sigma)$ can be chosen arbitrarily from the class of functional densities centered to the origin, provided that $\wp(x|\sigma)\in C^2(\R).$ For the both above-mentioned variants it holds
$$ \lim_{\sigma \rightarrow 0_+} \wp(x|\sigma) \stackrel{\mathscr{D'}}{=} \lim_{\sigma \rightarrow 0_+} B(x|\sigma) \stackrel{\mathscr{D'}}{=} \delta(x),$$
where $\delta(x)$ is the well-known Dirac delta pulse. By a transition to this zero-variance limit one can, in fact, obtain a discrete variant of the smoothed number of particles (\ref{SNP}).

\subsection{Summary of terminology used} \label{subsec:APP-02}

For comfort and better understanding, in the following table we summarize a terminology used in this paper.

 \begin{table}[h]
\caption{Terminology.}\label{Tab:terminology} \centering \footnotesize
  \vspace*{0.2cm}
\begin{tabular}{|l|l|l|}
  \hline
 Technical term &  Symbol  & Explanation \\
  \hline
  Vehicular Headway Modeling & VHM & scientific field dealing with inter-vehicular gaps\\
  detector line & & a static location of detector \\
  primary traffic quantities & & gauged by detector or stated by exact formulas\\
  secondary traffic quantities & & calculated from primary ones by approximations\\
phase (state) variables  & $\varrho,$ $I,$ $V$ & density, intensity, and mean speed\\
equation of continuity & $\frac{\p \varrho}{\p \tau}=- \frac{\p I}{\p \xi}$ & law of conservation for traffic flow\\
phase segment (segmentation zone) & $\Psi$ & small sub-area v ID plane\\
first phase relation & $\Omega_{\mathtt{ID}}$ & all existing intensity-density pairs \\
second phase relation & $\Omega_{\mathtt{VD}}$ & all existing velocity-density pairs \\
phase diagram & & surface graph of phase relations\\
time headway & $z$ & time between two cars as they pass a detector\\
time clearance & $t$ & time gap between two successive vehicles\\
distance headway & $s$ & space between two front bumpers in a fixed time\\
distance clearance & $r$& space gap between two successive vehicles\\
individual velocity & $v$& speed of a vehicle\\
scaled time clearance & $y$ & time clearance after the unification applied\\
scaled distance clearance & $x$ & space clearance after the unification applied\\
sample & & set of several succeeding vehicles \\
sample of time headways ($i$th) & $Z_i$ & set of time headways in one sample\\
sample of time clearances & $T_i$ & set of time clearances in one sample\\
sample of space headways & $S_i$ & set of space headways in one sample\\
sample of space clearances & $R_i$ & set of space gaps in one sample\\
sample of velocities & $\Upsilon_i$ & set of velocities related to a sample\\
sample of scaled time clearances & $Y_i$ & set of scaled time clearances in one sample\\
sample of scaled space gaps & $X_i$ & set of scaled space gaps in one sample\\
smoothed number of particles & $N(\xi,\tau)$ & number of cars, mathematically optimized\\
segmented index set & $F_\Psi$ & sample indices belonging to $\Psi$\\
segmented time clearances & $Y_\Psi$ & set of unified time clearances belonging to $\Psi$\\
segmented space clearances & $X_\Psi$ & set of unified space gaps belonging to $\Psi$\\
segmented velocities & $\Upsilon_\Psi$ & set of velocities belonging to $\Psi$\\
  \hline
\end{tabular}
\end{table}


\begin{thebibliography}{10}
\bibitem{Borsalino} Borsalino is the oldest Italian company specializing in the manufacture of luxury hats.
\end{thebibliography}
\end{document}